\documentclass[preprint,reprint,amsmath,showpacs,amssymb,superscriptaddress]{revtex4-1}
\usepackage{graphicx}
\usepackage{dcolumn}
\usepackage{bm}

\begin{document}
\title{Landau level spin diode in a GaAs two dimensional hole system}

\author{O. Klochan}
\email{klochan@phys.unsw.edu.au}
\author{A.R. Hamilton}
\affiliation{School of Physics, University of New South Wales,
Sydney NSW 2052, Australia.}

\author{K. das Gupta}
\affiliation{Cavendish Laboratory, University of Cambridge, Cambridge, CB3 0HE, U.K.}
\affiliation{Department of Physics, IIT Bombay, Mumbai 400076, India}	

\author{F. Sfigakis}
\author{H.E. Beere}
\author{ D.A.  Ritchie}
\affiliation{Cavendish Laboratory, University of Cambridge, Cambridge, CB3 0HE, U.K.}

\date{\today}

\begin{abstract}

We have fabricated and characterized the Landau level spin diode in GaAs two dimensional hole system. We used the hole Landau level spin diode to probe the hyperfine coupling between the hole and nuclear spins and found no detectable net nuclear polarization,  indicating that hole-nuclear spin flip-flop processes are suppressed by at least three  orders of magnitude compared to GaAs electron systems.

\end{abstract}

\maketitle

Studying the coupling between spins of charge carriers and nuclear spins is a rapidly expanding field due to possible applications for quantum information processing~\cite{ChekhovichNatMat13}. This coupling is a major source of electron spin decoherence in GaAs spin qubits, because both Ga and As  have nonzero nuclear spins. Several approaches have been developed to suppress this unwanted interaction~\cite{ChekhovichNatMat13, AwschalomScience13}, including NMR  techniques such as spin-echo, dynamic nuclear polarization as well as the  elimination of nuclear spins by using nuclear free materials such as $^{12}$C or $^{28}$Si. An alternative  approach to minimising spin decoherence is to use hole spins since hyperfine coupling is much weaker than for electrons~\cite{TartakovskiiNatPhot12}. The contact hyperfine interaction between an electron spin $S$ and nuclear spins $I$ is given by ~\cite{CoishPSS09}:  $H_{e,n}=\frac{1}{2}M_S\{I_+S_- + I_-S_+\} + M_SI_ZS_Z$, where  $M_S=\frac{2\mu_0}{3}\gamma_e\gamma_n$ . The first 'flip-flop' term describes a process where the electron spin changes its orientation, and the nuclear spin  simultaneously changes its  orientation in the opposite direction.  The second term describes  the influence of nuclear polarization on the electrons via the effective Overhauser magnetic field. Valence band holes are formed by $p$-type atomic orbitals and so have zero overlap with the nuclei, unlike conduction band electrons which come from $s$-type orbitals. This eliminates the  contact hyperfine interaction leaving only the weaker dipolar hyperfine interaction. For light holes (LH) the nuclear spin coupling takes a similar form to that for electrons~\cite{TestelinPRB09}:  $H_{LH,n}=\frac{M_P}{3}[(I_+S_- + I_-S_+) + I_ZS_Z]$, where $M_P=\frac{3}{8\pi}M_S$ , but  $M_P$ is almost an order of magnitude smaller than $M_S$~\cite{FischerPRB08}. For pure heavy holes (HH) there is no spin-flip flop term~\cite{TestelinPRB09}, with only the Overhauser term present: $H_{HH,n}=\frac{M_P}{3}I_ZS_Z$.

Although there have been some electrical studies of the hyperfine coupling for holes in GaAs~\cite{KeaneNano11}, most studies  have probed holes optically in self-assembled quantum dots~\cite{ChekhovichPRL11, FallahiPRL10}. The optical  studies have confirmed theoretical predictions that the hyperfine coupling for holes is approximately an order of magnitude smaller compared to electrons, and that unlike electrons  the hole spin coherence time is  not limited by nuclear spin fluctuations~\cite{GreveNatPhys11, GreilichNatPhot11}. However, there have not been any studies of hole-nuclear spin coupling in gate defined quantum dots. Such studies are challenging because  the large hole effective mass in GaAs  $m_h = 0.2-0.5\times m_0$ makes it difficult to access the single hole regime in quantum dots. Previously we compared electron-nuclear and hole-nuclear spin coupling through   the breakdown of the quantum Hall effect in a narrow one-dimensional constriction~\cite{KeaneNano11}. However, this method relies on detecting small changes of the sample resistance,  $<1\%$. A much more sensitive technique to detect nuclear spin effects electrically is the Landau level diode technique, which directly measures tunnelling between spin resolved Landau levels~\cite{KanePRL88,KanePRB92,DixonPRB97}. Using this approach changes in the sample resistance can exceed $100\%$, which is two orders of magnitude more sensitive compared to previous studies~\cite{CorcolesPRB09,KeaneNano11}. Additionally the Landau level diode technique allows energy resolved inter Landau level tunneling to be studied, providing unique spectroscopic information. In this work we use Landau level diode technique to probe coupling of nuclear and hole spins in a GaAs two dimensional (2D) hole system in the quantum Hall regime at filling factor $\nu=2$. We separately contact the two lowest edge channels  and force holes to scatter between adjacent Landau levels with opposite spins of  $\pm3/2$. Similar measurements in electron systems (with spin $\pm1/2$) revealed a build up of nuclear spin polarization due to electron-nuclear spin flip-flop processes~\cite{KanePRB92,DixonPRB97}. However, in the present study we observe no detectable net nuclear polarization, indicating that hole-nuclear spin flip-flop processes are suppressed by  at least three  orders of magnitude compared to GaAs electron systems.

Our device is fabricated from an  undoped GaAs heterostructure, where holes are induced by a negative bias voltage $V_{TG}$ applied to an overall top gate. The active region of the heterostructure, grown by molecular beam epitaxy  on a (100) GaAs substrate, comprises of a  1~$\mu$m GaAs buffer layer  followed by 300~nm of undoped AlGaAs and capped by~10 nm of GaAs. Electrical contact to the 2D hole system was achieved using annealed AuBe ohmic contacts.  Standard electron beam lithography, metal deposition and lift-off were used to form depletion surface gates. A 500~nm thick layer of polyimide was used as an insulator to isolate the overall metal top gate from the surface gates and the ohmic contacts following the procedure developed in Ref.~\cite{HarrellAPL99}.   We characterize our sample using low field  Hall measurements to find the  2D hole density $p$ in the range  between $9\times10^{10}$ and $1.44\times10^{11}$~cm$^{-2}$ ($V_{TG}=-4$ to $-7$~V) and the hole mobility  $\mu$  approximately constant at $300 000$~cm$^{2}$/Vs. Experiments were performed  in a dilution fridge with a base temperature of $25$~mK using DC measurements.
\begin{figure}
\includegraphics[width = 8cm]{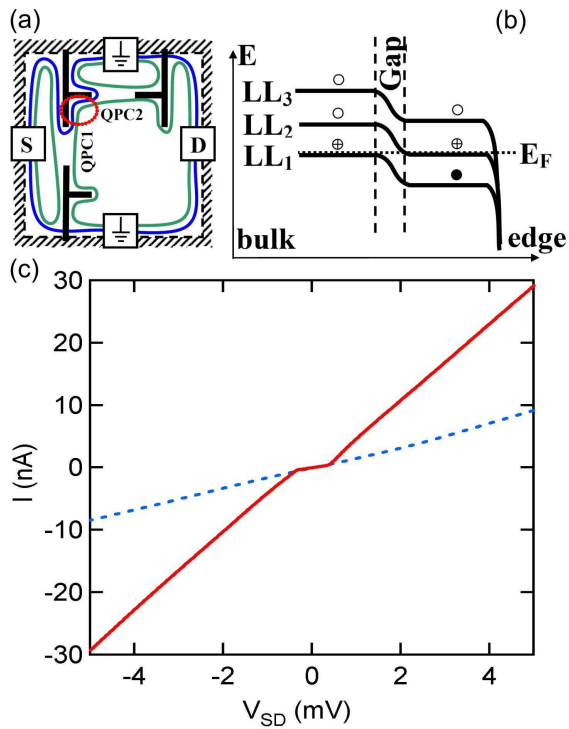}
\caption{\label{fig1} (Color online) (a) Schematic of the  device and measurement setup. The magnetic field  $B=\pm3$~T ensures  $\nu=2$ in the 2D hole system ($V_{TG}=-7$~V, $p = 1.44 \times 10^{11}$~cm$^{-2}$). The T-shaped black lines indicate surface gates which direct the edge channels and define QPC1 and QPC2.  The solid blue and green concentric lines show the edge channels LL1 and LL2. The T-shaped gates are biased such that only LL1 is transmitted through  QPC1 and QPC2. The red dotted circle indicates the $1.4~\mu$m long interaction region; (b) Schematic energy diagram of the Landau levels (LL) in the interaction region at zero source-drain bias. The occupation of the Landau levels is shown by the circles: solid = occupied; open = empty, crossed = partially occupied; (c) IV characteristics measured for edge channels running clockwise (red solid line, $B=-3$~T) and counterclockwise (blue dashed line, $B=3$~T).  }
\end{figure} 

A schematic of our device is shown in Fig.~1(a). The mesa edge is shown by the dashed square. There are four ohmic contacts located at the edges of the mesa: source (S), drain (D) and two connected to ground.  Three T-shaped surface gates run across the mesa edge  and form two short constrictions (quantum point contacts): QPC1 - separating S and D contacts, and QPC2  - separating top grounded  and D contacts. Both the 2D hole density and magnetic field are tuned such that $\nu=2$ in the bulk and two edge channels (blue and green concentric lines) run clockwise along the mesa edge. The T-shaped surface gates are biased such that only one of the edge channels goes through the constrictions and the other one is reflected ($\nu=1$ in QPC1 and QPC2). This allows separate electrical contact to the edge channels and direct I-V measurements to be performed between the two edge channels. This technique uses a property of the  Landau levels  near the edge of the sample, where they  flatten out to form  edge states~\cite{ChklovskiiPRB92}. As shown in Fig.~1(a), in this configuration the source essentially contacts the lowest Landau level (LL1, blue edge channel), whereas the drain contacts the second Landau level (LL2, green edge channel). Current can only flow from source to drain if holes can transfer from LL1 to LL2 in the small overlap region ($\approx 1.4~\mu$m long) highlighted by the red dotted circle. In this region the edge state structure is similar to that of a diode, as shown in Fig.~1(b) With no bias applied there is no transport across the two edge channels, so we expect zero current. When we apply a positive (forward) bias, above a certain threshold voltage $V_{TH}$,  the levels  align causing a rapid increase in current.  Applying a negative (reverse) bias  should not allow current until the onset of Zener tunnelling through the gap.  Therefore the IV characteristics of such a device will look like that of a diode.  
In the case of electrons  at $\nu=2$ the two occupied edge channels are spin polarized, so when there is a current flowing one  Landau level to the other, the electron spin has to flip, which can cause nuclear spin to flop and thus build up a net nuclear polarization over time. This net nuclear polarization acts back on electrons and changes their Zeeman energy (via the Overhauser field) leading to hysteresis and characteristic long time dependence in the IV characteristics as the nuclear spin polarization  decays very slowly~\cite{KanePRB92,WaldPRL94,DixonPRB97}. 

The red trace in Fig.~1(c) is a typical IV trace measured in our device. We observe a strong nonlinear behaviour confirming that our device operates as a  Landau level diode \cite{KanePRL88,DixonPRB97}.  At low bias ($-0.37 < V_{SD} < 0.42$~mV) only a very small current flows through the  device (we call this the suppressed transport region).  As we increase the voltage past the suppressed transport region a large current starts to flow between the two edge channels. We note that the current in the  suppressed transport region is not quite zero but has a linear I-V dependence corresponding to a resistance exceeding $500$~kOhm. This is  due to the finite resistance of the  ohmic contacts~\cite{HaugSST93} -  essentially, not all the current flows to  the top grounded  contact, allowing a small leakage current via LL1. This is confirmed by reversing $B$, so that the edge channels  run anticlockwise.  In this case  there should be no current registered at drain, since both LL1 and LL2 go directly to  the bottom grounded ohmic contact. However there is a finite source-drain current as  shown by the blue dashed line in Fig.~1(c, which has  the same resistance as in  the suppressed transport region (red trace, low $V_{SD}$).

The diode-like behaviour of the I-V characteristics in Fig. 1(c) are almost symmetric, in contrast to similar data for electrons in GaAs~\cite{DixonPRB97,WurtzPRB02}. This may be due to proximity of higher (empty) Landau levels, due to the large hole mass and larger hole g-factor. In electron systems~\cite{DixonPRB97,WurtzPRB02}, the asymmetry of the I-V characteristics is due to large energy mismatch between the Zeeman energy $E_Z = \Delta E_{1,2}$ and the cyclotron energy $ \hbar\omega_c = \Delta E_{2,3}$ such that $\hbar\omega_c >> E_Z$. If  the electron Landau diode is operated in a regime with higher filling factor in the bulk (e.g. $\nu=4$ in the bulk and $\nu=2$ in the QPCs), the energy scales are comparable and  the IV traces look more symmetric~\cite{WurtzPRB02, DeviatovPrivate}. 

The Landau level diode technique allows us to perform spectroscopy of the  energy separation between the lowest two hole Landau levels in magnetic field. We extract this energy gap  by measuring   the positive threshold voltage ($eV_{TH}=\Delta E_{1,2}$), which corresponds to the situation when the Landau levels are aligned~\cite{KanePRL88,KanePRB92,DixonPRB97}.   Figure~2(a) shows a series of IV traces measured for different $V_{TG}$ (carrier density) and corresponding $B$  so that  both the bulk filling factor is maintained at $\nu=2$ and  $\nu=1$ is maintained in  QPC1 and QPC2 for all traces. As we reduce the density from $p = 1.44\times10^{11}$~cm$^{-2}$ (violet trace) to $p = 9\times10^{10}$~cm$^{-2}$ (red trace) the width of the suppressed transport region gets smaller, indicating the reduction of  $\Delta E_{1,2}$ with decreasing $B$. For electrons systems,  the forward bias threshold voltage is determined by the Zeeman energy and evolves linearly in $B$. In contrast for hole systems it is well known that   the Landau levels evolve non-linearly in $B$~\cite{WinklerBook}. In Fig.~2(b) we plot the extracted  threshold voltages~\cite{Vth} ($V_{th}$) as a function of $B$. Surprisingly, for the positive threshold we observe an almost  linear dependence on $B$. To compare our results to  electron systems, from a linear fit to the data we extract the splitting rate, to obtain  effective  $g$-factor ($g^*$). We  find $g^*=3.25 \pm 0.15$, significantly exceeding values measured in electron Landau diodes~\cite{WurtzPRB02}, although somewhat less than the ideal low-field limit of $g^*=7.2$ for heavy holes. 
\begin{figure}
\includegraphics[width = 8cm]{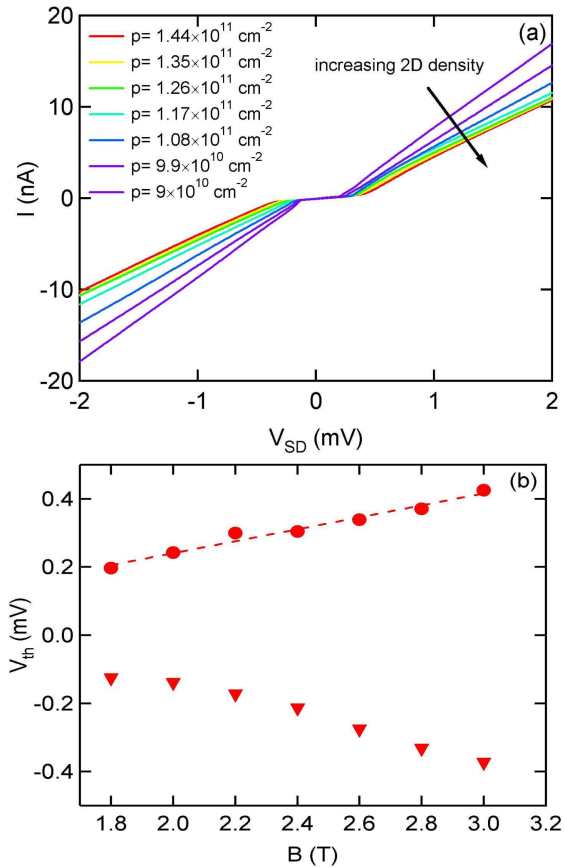}
\caption{\label{fig3} (Color online)(a) IV traces for different $V_{TG}$ between $-4$ and  $-7$~V in steps of $0.5$~V  corresponding to a  density range $0.9 < p < 1.44 \times 10^{11}$~cm$^{-2}$ and magnetic field range $1.8 <B < 3$~T. (b) Positive (circles) and negative (triangles) threshold voltages $V_{th}$ extracted from Fig.~3(a) plotted vs $B$. The dashed line indicate the best linear fit through the data points (circles). }
\end{figure}

Having demonstrated the operation of the hole Landau diode we use it to probe  hyperfine coupling between hole and nuclear spins. The two characteristic signatures of hyperfine coupling observed in electron systems are: 1)~hysteretic I-V traces and 2)~time dependence of source-drain current on the time scales of tens of seconds~\cite{KanePRB92,WaldPRL94,DixonPRB97,WurtzPRB02, KeaneNano11}. In electron Landau diodes, when a reverse bias is applied above the threshold voltage, electrons  tunnel between  spin polarized Landau levels with different spin orientations, so the electron spin has to flip. The excess spin is transferred to the nuclei via the hyperfine interaction. These electron-nuclear spin flip-flop processes build up a net nuclear polarization over time. This polarization is detected as time dependent  I-V characteristics via the effect of the nuclear Overhauser field on the electron Zeeman energy. For holes the situation is very different as the nature of the lowest two Landau levels is not simple spin $\pm 1/2$; in general hole Landau levels are mixtures of heavy- and light-hole states~\cite{WinklerBook}. Theoretical calculations for holes suggest that  the lowest Landau level is a pure HH state  with total angular momentum $j=-3/2$~\cite{WinklerBook}. The second Landau level is mixed and has $j=3/2$. In the Landau level diode, when the current flows between these two levels, the total angular momentum  has to change by $3\hbar$, which is not a very efficient process.  

To look for the hysteresis we sweep $V_{SD}$ back and forth. If the sweep rate is comparable to the nuclear spin relaxation time ($\sim 10$~s), the I-V traces will show hysteresis. Previous studies of electrically detected electron-nuclear spin coupling typically show  $<1\%$ hysteresis due to nuclear spins~\cite{CorcolesPRB09,KeaneNano11}. In contrast, in similar studies that used the Landau level technique~\cite{DixonPRB97} the magnitude of the hysteresis can exceed  $100\%$ allowing much more sensitive detection of nuclear polarization. Figure~3(a) shows three consecutive IV traces  taken with a  sweep rate  of  25, 60 and 300 seconds per trace. All traces sit on top of each over with no sign of hysteresis. To increase the sensitivity of our measurement, we attempt to detect time dependent current relaxation at several  bias voltages above $V_{th}$ on both positive and negative sides of the IV trace as shown in Fig.~3(b). First, we set the  bias voltage to zero and dwell for 5 min so that no current flows between the Landau levels and the nuclear spins are in equilibrium (unpolarized). Then we rapidly set $V_{SD}$  to the operating point exceeding the threshold voltage, to  initiate inter-Landau levels transitions and build up a nuclear spin polarization (providing the spin flip-flop mechanism is available). Finally, we monitor the current   as a function of time  to  detect any change in nuclear polarization through the Overhauser field.  As shown in Fig.~3(b) we observe no sign of current relaxation down to the noise level in our measurement setup  ($0.3\%$). To reduce the noise even further we averaged  10 traces to get down to $0.1\%$ signal-to-noise ratio  and still there is no sign of current relaxation.  This is approximately three orders of magnitude less than the typical magnitude of the current relaxation and bias hysteresis  measured in electron systems~\cite{DixonPRB97}. 
\begin{figure}
\includegraphics[width = 8cm]{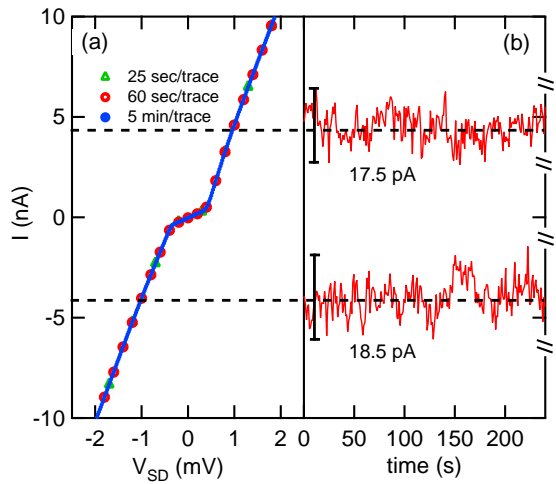}
\caption{\label{fig4} (Color online) (a) Three IV traces measured  with a different sweep rate ranging from 25 sec/trace (greed open triangles), 60 sec/trace (red open circles) and 5 min/trace (blue solid circles). (b) Two time traces taken at $V_SD=\pm 1$~mV for 4 min  with a dwell of 5~min at $V_{SD}= 0$~V in between the time traces. The scale bars indicate the magnitude of current oscillations for a single trace with no averaging.}
\end{figure}

Recent optical studies in self-assembled quantum dots have shown that the hyperfine coupling for holes is at least 10 times weaker than for electrons~\cite{ChekhovichPRL11,FallahiPRL10}.  Taking this results into account, our data  indicate that in addition to suppressed sensitivity to the Overhauser field, the  hole-nuclear spin flip-flop processes are at least three orders of magnitude less efficient at generating nuclear polarization. However, Ref.~\cite{ChekhovichNatPhys13} indicate significant mixing of the $d$-shell atomic orbitals into the $p$-like hole wavefunction. This mixing allows spin flips between heavy hole states, in contrast to the case of pure p-shell heavy holes for which the hyperfine coupling has an Ising form~\cite{FischerPRB08}. Thus, there are two possible mechanisms for heavy hole-nuclear spin flip-flop processes: LH-HH mixing or mixing of $p$- and $d$- atomic orbitals for pure HH. In either case, inter-Landau level transitions should be able to cause a nuclear spin polarization, which will cause detectable Overhauser field. The fact that we do not observe any nuclear polarization is therefore intriguing and suggests that additional work is necessary to understand nuclear spin coupling in GaAs hole systems and the role of spin-orbit interaction.  		

To summarize we have fabricated and characterized a Landau level spin diode in GaAs two dimensional hole system. We used the hole Landau level spin diode to show that surprisingly the splitting on the lowest two Landau levels is linear in $B$ over the range of our measurements.  We have observed no evidence of  hyperfine coupling, indicating that hole-nuclear spin flip-flop processes are suppressed by at least three orders of magnitude compared to  GaAs  electron systems.

This work was funded by the Australian Research Council through the Discovery Projects Program  and by the Australian
Government under the Australia-India Strategic Research Fund. We thank U. Z\"{u}licke, R. Winkler, E. V. Deviatov, V. T. Dolgopolov  for illuminating discussions. Experimental devices for this study were fabricated using the Australian National Fabrication Facility, UNSW.


\begin{thebibliography}{99}
\bibitem{ChekhovichNatMat13} E. A. Chekhovich  {\it et al.}, Nature Mat. {\bf 12}, 494 (2013).
\bibitem{AwschalomScience13} D. D. Awschalom {\it et al.}, Science {\bf 339}, 1174 (2013). 
\bibitem{TartakovskiiNatPhot12} A. Tartakovskii, Nature Phot. {\bf 5}, 647 (2011). 
\bibitem{CoishPSS09} W. A. Coish, J. Baugh, Nuclear Spins in Nanostructures, Phys. Stat. Sol. (b){\bf 246}, 2203 (2009).
\bibitem{TestelinPRB09} C. Testelin {\it et al.} , Phys. Rev. B {\bf 79}, 195440 (2009).
\bibitem{FischerPRB08} J. Fischer  {\it et al.}, Phys. Rev. B {\bf 78}, 155329 (2008).
\bibitem{KeaneNano11} Z. K. Keane  {\it et al.} Nano Lett. {\bf 11}, 3147 (2011).
\bibitem{ChekhovichPRL11} E. A. Chekhovich {\it et al.}, Phys. Rev. Lett. {\bf 106}, 027402 (2011).
\bibitem{FallahiPRL10} P. Fallahi {\it et al.}, Phys. Rev. Lett. {\bf 105}, 257402 (2010).
\bibitem{GreveNatPhys11} K. De Greve {\it et al.}, Nature Phys. {\bf 7}, 872, (2011).
\bibitem{GreilichNatPhot11} A. Greilich {\it et al.}, Nature Phot. {\bf 5}, 702, (2011).
\bibitem{KanePRL88} B. E. Kane {\it et al.}, Phys. Rev. Lett. {\bf 61}, 1123 (1988).
\bibitem{KanePRB92} B. E. Kane  {\it et al.}, Phys. Rev. B {\bf 46}, 7264 (1992).
\bibitem{DixonPRB97} D. C. Dixon  {\it et al.}, Phys. Rev. B {\bf 56}, 4743 (1997).
\bibitem{CorcolesPRB09} A. Corcoles  {\it et al.}, Phys. Rev. B {\bf 80}, 115326 (2009).
\bibitem{HarrellAPL99} R. H. Harrell  {\it et al.}, Appl. Phys. Lett.  {\bf 74}, 2328 (1999).
\bibitem{ChklovskiiPRB92} D. B. Chklovskii  {\it et al.}, Phys. Rev. B {\bf 46}, 4026 (1992).
\bibitem{WaldPRL94} K. R. Wald {\it et al.}, Phys. Rev. Lett. {\bf 73}, 1011 (1994).
\bibitem{HaugSST93} R. J. Haug, Semicond. Sci. Technol.  {\bf 8}, 131 (1993).
\bibitem{WurtzPRB02} A. W\"{u}rtz  {\it et al.}, Phys. Rev. B {\bf 65}, 075303 (2002).
\bibitem{DeviatovPrivate} E. V. Deviatov,  {\it Private Communication}.
\bibitem{WinklerBook} R. Winkler, "Spin-orbit coupling effects in two-dimensional electron and hole systems", Springer-Verlag, 2003.
\bibitem{Vth} Similarly to Ref.~\cite{DixonPRB97} the threshold voltage $V_{th}$ is defined as an intersect of two  lines  fitted to the region where almost no current flows through the device and where large current flows through the device.
\bibitem{ChekhovichNatPhys13} E. A. Chekhovich  {\it et al.}, Nature Phys. {\bf 9}, 74 (2013).






\end{thebibliography}
\end{document}